\documentclass[prl,aps,showpacs,superscriptaddress,floatfix,twocolumn]{revtex4}
\usepackage{graphicx}
\begin{document}

\title{Light-matter excitations in the ultra-strong coupling regime}
\author{Aji A. Anappara}
\affiliation{NEST CNR-INFM and Scuola Normale Superiore, Piazza dei Cavalieri 7, 
I-56126 Pisa (Italy)}
\author{Simone De Liberato}
\affiliation{Laboratoire ÒMat\'eriaux et Ph\'enom\`enes QuantiquesÓ, Universit\'e Paris Diderot - Paris 7 and CNRS, UMR 7162, B\^atiment Condorcet, 75013 Paris (France)} 
\affiliation{Laboratoire Pierre Aigrain, Ecole Normale Sup\'erieure and CNRS, UMR 8551, 75005 Paris (France)}
\author{Alessandro Tredicucci} 
\email{a.tredicucci@sns.it}
\affiliation{NEST CNR-INFM and Scuola Normale Superiore, Piazza dei Cavalieri 7, 
I-56126 Pisa (Italy)} 
\author{Cristiano Ciuti}
\affiliation{Laboratoire ÒMat\'eriaux et Ph\'enom\`enes QuantiquesÓ, Universit\'e Paris Diderot - Paris 7 and CNRS, UMR 7162, B\^atiment Condorcet, 75013 Paris (France)} 
\author{Giorgio Biasiol}
\affiliation{Laboratorio Nazionale TASC CNR-INFM, Area Science Park, SS 14 Km 163.5, Basovizza, I-34012 Trieste (Italy)}
\author{Lucia Sorba} 
\affiliation{NEST CNR-INFM and Scuola Normale Superiore, Piazza dei Cavalieri 7, 
I-56126 Pisa (Italy)}
\author{Fabio Beltram}
\affiliation{NEST CNR-INFM and Scuola Normale Superiore, Piazza dei Cavalieri 7, 
I-56126 Pisa (Italy)}

\date{\today}

\begin{abstract}
In a microcavity, light-matter coupling is quantified by the vacuum Rabi frequency $\Omega_R$. When $\Omega_R$ is larger than radiative and non-radiative loss rates, the system eigenstates (polaritons) are linear superposition of photonic and electronic excitations, a condition actively investigated in diverse physical implementations. Recently, a quantum electrodynamic regime (ultra-strong coupling) was predicted when $\Omega_R$ becomes comparable to the transition frequency. Here we report unambiguous signatures of this regime in a quantum-well intersubband microcavity. Measuring the cavity-polariton dispersion in a room-temperature linear optical experiment, we directly observe the anti-resonant light-matter coupling and the photon-energy renormalization of the vacuum field.
\end{abstract}

\pacs{71.36.+c, 78.67.De, 42.50.Pq, 73.21.Fg}

\maketitle

The strong-coupling  regime between a dipole-allowed electronic transition and the photonic mode of a microcavity manifests itself in the lifting of the degeneracy between the two modes, with an anti-crossing behaviour of the new polariton eigenstates, separated by an energy  termed vacuum-Rabi splitting (VRS) in atomic physics~\cite{8}, or cavity-polariton splitting in solid-state systems~\cite{2}. This regime is actively investigated in many research fields, such as ultracold atoms in optical cavities~\cite{3}, Cooper-pair boxes in microwave resonators \cite{9}, excitonic transitions in semiconductor microcavities \cite{5} and surface-plasmon resonators~\cite{6}.

The magnitude of light-matter coupling in atomic systems is limited by the intrinsically small dipole moment of the transitions. Typical values for a single atom are $\Omega_R\approx10^{-7}-10^{-6}$~$\omega_{12}$, $\omega_{12}$ being the transition frequency~\cite{8}. Circuit quantum electrodynamics in superconducting systems, instead, can generally produce much larger $\Omega_R/\omega_{12}$ ratios, of the order of few percent \cite{9}. Even larger values are possible using intersubband transitions between two-dimensional electronic states within the conduction band of semiconductor heterostructures~\cite{11}.

With increasing $\Omega_R/\omega_{12}$, terms of the interaction Hamiltonian that are otherwise negligible become more and more relevant. This leads to profound modifications in the very nature of the quantum states of the system. These changes stem from the renormalization of the electromagnetic field and non-resonant contributions, effects one intuitively associates only to strongly-driven systems and not to vacuum-field interaction. The energy of the excitations is affected and a new ÒsqueezedÓ ground state is defined containing a finite non-zero number of virtual photons. Theoretical investigations reveal that these virtual photons can be released in correlated pairs by non-adiabatic manipulation of the light-matter coupling, a phenomenon reminiscent of the dynamical Casimir effect~\cite{12,13}. These peculiar phenomena prompted researchers to coin the term \textit{ultra-strong coupling} to identify this condition.

In this report we demonstrate a semiconductor microcavity displaying unambiguous signatures of the ultra-strong coupling regime of light-matter interaction, even at room temperature. The structure is based on intersubband transitions, which, beyond the large coupling strength, also offer ample possibilities for its external control \cite{14, 15}.

Intersubband transitions involve levels originating from the quantum-mechanical confinement of charge carriers in one direction. Energy, carrier density, and matrix elements are the relevant parameters of the resonance, and can be tailored through structural design. The strong coupling with the electromagnetic mode of a planar semiconductor resonator, and the corresponding formation of \textit{intersubband polaritons}, was observed in GaAs/AlGaAs \cite{14,16} and InAs/AlSb \cite{17} material systems, up to room temperature. These solid-state systems can be grown by mature epitaxial growth techniques such as molecular beam epitaxy (MBE) and represent optimal candidates to realize the ultra-strong coupling regime of light-matter interaction.
 
The Hamiltonian of the intersubband microcavity can be written using a bosonic approximation, since the excitation density of the transition (intersubband excitations per unit area of the sample) is very small compared to the density of the two-dimensional electron gas \cite{12}. The Hopfield-like Hamiltonian then takes the form:
\begin{equation}
\label{eq1}
H=H_{res}+H_{dia}+H_{anti-res}.
\end{equation}
It consists of three qualitatively different contributions that correspond to the three main terms of the electromagnetic interaction. $H_{res}$ is given by: 
\begin{widetext}
\begin{equation}
\label{eq2}
H_{res}=\hbar\sum_k\left[\omega_{cav}(k)\left(a^\dagger_ka_k+\frac{1}{2}\right)+\omega_{12}b^\dagger_kb_k+i\Omega_{R,k}\left(a^\dagger_kb_k-a_kb^\dagger_k\right)\right],
\end{equation}
\end{widetext}
where $a^\dagger_k$ ($a_k$) is the creation (annihilation) operator for the fundamental cavity-photon mode with in-plane wavevector $k$ and frequency $\omega_{cav}(k)$, $\Omega_{R,k}$ is the $k$-dependent Rabi coupling frequency, and $b^\dagger_k$ ($b_k$) is the creation (annihilation) operator of the bright intersubband-excitation mode of the doped multiple quantum well structure. $H_{res}$ describes the energy of the bare cavity photon, the intersubband polarization field, and the resonant part of the light-matter interaction (corresponding to the creation / annihilation of one photon with the concomitant annihilation / creation of an intersubband excitation with the same in-plane wavevector). 

The middle contribution in Eq.~(\ref{eq1}) contains the diamagnetic term (proportional to the square of the vector potential A) and gives a renormalization of the photon energy due to the interaction with matter:
\begin{equation}
\label{eq3}
H_{dia}=\hbar\sum_kD_k\left(a^\dagger_ka_k+a_ka^\dagger_k\right),
\end{equation}
where, for a quantum well, the diamagnetic coupling constant $D_k$ is approximately given by $D_k\approx\Omega^2_{R,k}/\omega_{12}$ \cite{12}.

The last contribution in Eq.~(\ref{eq1}) is represented by the so-called Òanti-resonantÓ terms, corresponding to the simultaneous creation and annihilation of two excitations with opposite in-plane wavevectors:
\begin{widetext}
\begin{equation}
\label{eq4}
H_{anti-res}=\hbar\sum_k\left[i\Omega_{R,k}\left(a_kb_{-k}-a^\dagger_kb^\dagger_{-k}\right)+D_k\left(a_ka_{-k}+a^\dagger_ka^\dagger_{-k}\right)\right].
\end{equation}
\end{widetext}
Matrix elements of Eq.~(\ref{eq4}) are non-zero only when coupling states with different total number of cavity photons and intersubband excitations. This term is suppressed in first order perturbation theory. Neglecting $H_{anti-res}$, the Hamiltonian (\ref{eq1}) commutes with the boson number and can be block-diagonalized in a finite dimension subspace. This kind of approximation is the keystone of all analytical results in the field of light-matter interaction, and is usually known as rotating wave approximation (RWA)~\cite{19}. Normally, it is violated only in the case of dressed states in strongly driven systems with a large number of photons; experimental evidence stemming from the observation of energy shifts or forbidden transitions~\cite{19, 20}.

\begin{figure}[t!]
\begin{center}
\includegraphics*[width=0.5\textwidth]{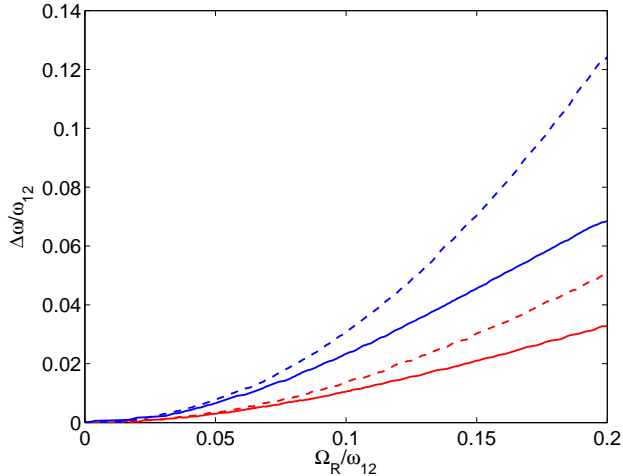}
\caption{Difference from the full-Hamiltonian eigenvalues of the polariton energies calculated either without the $H_{anti-res}$ term (red) or without both $H_{anti-res}$ and $H_{dia}$ (blue) plotted as a function of coupling strength (solid and dashed lines refer to the lower and upper polariton branch, respectively). This calculation was performed considering a fixed resonant angle of 60$^\circ$. As one can see, deviations amount to $\sim5$ \% 
already for $\Omega_R/\omega_{12}\sim 0.1$.}
\label{fig:1}
\end{center}
\end{figure}

The $H_{dia}$ and $H_{anti-res}$ contributions to the interaction with the vacuum-field have long been elusive, owing to the fact that the process involves states with zero photon number, and, as such, these terms are usually negligible. Yet they represent the hallmark of the ultra-strong-coupling regime and are at the origin of the peculiar quantum nature of the states. A simple spectroscopic identification (e.g. based on the excitation energies) would be impossible in most microcavity systems. On the other hand, the situation of intersubband microcavities, which use a planar geometry with a resonator designed to operate at oblique incidence, is quite special. Measurements at large angles, in fact, highlight energy deviations of the polariton dispersion, which can easily become of the order of several percent (see Fig.~\ref{fig:1}). Furthermore the dispersion of the uncoupled modes can be separately measured to allow a fair comparison with theoretical models.

\begin{figure}[t]
\begin{center}
\includegraphics*[width=0.5\textwidth]{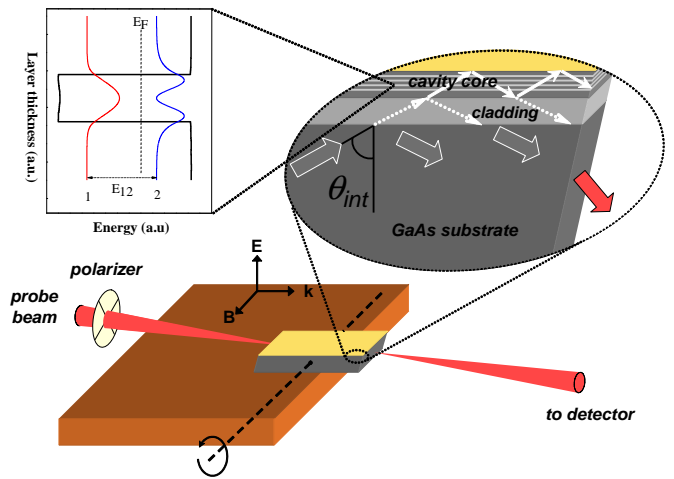}
\caption{Scheme of the experimental set-up employed for the angle-resolved reflectance measurements. The prism-shaped sample is mounted on a copper block and can be rotated to vary the internal incidence angle. The blown up detail of the waveguide illustrates the working principle of the resonator. The band profile and squared moduli of the subband envelope functions of one of the quantum wells (calculated solving self-consistently the Poisson-Schr\"odinger equation) are shown in the top left diagram.
}
\label{fig:2}
\end{center}
\end{figure}

\begin{figure}[t]
\begin{center}
\includegraphics*[width=0.5\textwidth]{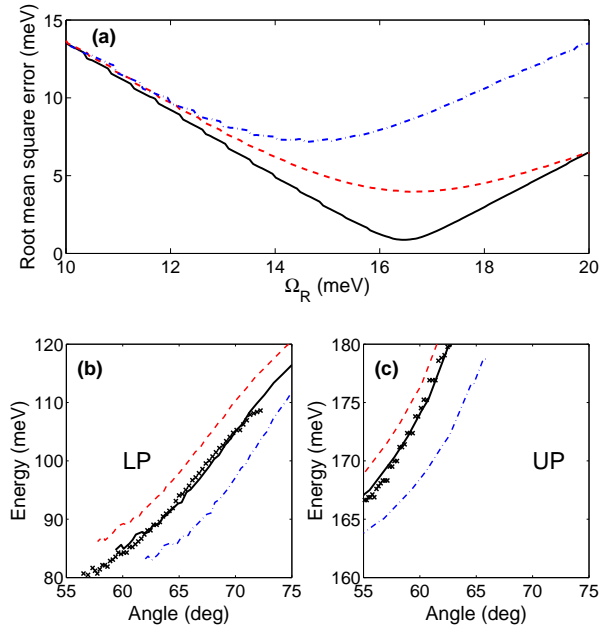}
\caption{Panel (a): root mean square deviation from measured dispersion of the calculated polariton energies as a function of the vacuum Rabi energy, the only fitting parameter. The red curve refers to the case of no $H_{anti-res}$ terms, while the blue to the case without both $H_{dia}$ and $H_{anti-res}$ contributions. The black line is for the full Hamiltonian. Bottom panels: angular dispersions of the lower (b) and upper (c) polaritons in the three cases (full Hamiltonian for the black solid line, without $H_{anti-res}$ for the dashed red line and without both $H_{dia}$ and $H_{anti-res}$ for the dash-dotted blue line), compared to experimental data (black crosses). The $\Omega_R$ used are the ones that minimize the root mean square deviation in the panel (a).
}
\label{fig:3}
\end{center}
\end{figure}

Optical confinement in the microcavity used in this investigation is based at the bottom end on the total-internal reflection from a low refractive-index cladding and at the top on the reflection from a semiconductor-metal interface (Fig.~\ref{fig:2}). The heterostructure was grown by solid-source MBE on an undoped GaAs (001) substrate \cite{epaps}. The cladding region was realized by sandwiching a 1.65 $\mu$m AlAs layer between two GaAs layers doped to $5\times10^{18}$~cm$^{-3}$, each having a thickness of 150 nm. The active region consists of 70 repeats of 
n-doped 6.5 nm thick GaAs quantum wells separated by 8 nm thick Al$_{0.35}$Ga$_{0.65}$As barriers.  Layer thickness was chosen so as to have only two bound subbands and ensure quantum de-coupling of adjacent wells. Doping level ($3.25\times10^{12}$~cm$^{-2}$ in each well) leads to population of the ground state only and to a single intersubband transition. 
Light was coupled into the microcavity through the substrate, and the cavity throughput was probed by angle-resolved reflectance measurements. The experimental geometry is detailed in Fig.~\ref{fig:2}: the sample was mechanically lapped into a wedge-shaped prism with the polished facets at an angle of 70$^\circ$ with respect to the cavity plane. The prism was mounted inside a Fourier-transform infrared spectrometer (FTIR), equipped with a cooled HgCdTe detector. A metallic wire-grid polarizer was inserted in the optical path to select the TM polarization of the probe beam. By manually rotating the sample holder, the angle between the infrared beam and the prism facet could be varied, enabling us to change the incident angle ($\theta_{int}$) on the cavity surface around the central value of 70$^\circ$ defined by the prism shape.
In the bottom panels of Fig. \ref{fig:3} are visible the angle-resolved peaks of the TM reflectance spectra for lower (b) and upper (c) polaritons in the frequency range of the intersubband transition as function of $\theta_{int}$. The spectra were collected at room temperature, with a resolution of 0.25~meV. The minimum splitting between the polariton dips is about 90 meV, $\sim59$ \% of the transition energy, although a precise anti-crossing point cannot be defined, this stems from the fact that polariton peaks at the same internal angle do not correspond to the same $k$ \cite{21}.

The bare intersubband-transition energy of the active region was measured in another wedge shaped prism, polished at 45$^\circ$ angle. The reflectance spectrum was collected at an internal angle of about 37$^\circ$, which excluded any cavity-induced shift of the intersubband absorption. The recorded transition energy is 152 meV, with a full width at half maximum (FWHM) of about 12 meV \cite{epaps}.

Since the bottom mirror utilizes total internal reflection, one cannot determine precisely the cavity resonance energy through measurements at zero incidence angle, where the intersubband transition does not couple to the radiation. We decided then to use a second sample, identical in the growth sequence, but without any doping in the active region in order to determine the energy dispersion of the cavity mode. The shift of the cavity refractive index induced by the absence of doping in the quantum wells was computed to be at most $\sim1$ \%, owing to the TM polarization of the light and large propagation angle.    The quality of the growth and the thickness difference between the two samples were checked using X-ray diffraction (XRD). No deviations were found within the XRD resolution of less than one percent \cite{epaps}. The reference sample was also wedged at an angle of 70$^\circ$ and the cavity dispersion determined from angle-resolved reflectance measurements \cite{epaps}. Scanning electron microscope (SEM) images of the cleaved facets of the two wedge-shaped samples were recorded to check the angular difference between the mechanically polished facets  \cite{epaps}. The deviation between the two samples was about 0.1$^\circ$, which does not cause a significant shift of the polariton peaks. 

Having determined experimentally the intersubband transition energy and the angle-dependent cavity mode frequency, we can  fit the data with the polariton dispersions calculated respectively using the full Hamiltonian of Eq. \ref{eq1}, the Hamiltonian without the anti-resonant terms, and the Hamiltonian in Eq. \ref{eq2}, that is the Hamiltonian without both the anti-resonant terms and the diamagnetic terms.
The only free fitting parameter in the three cases is the vacuum Rabi energy $\hbar\Omega_R$.
We calculated the root mean square deviation from the experimental data in the three cases and thus found the respective optimal fitting vacuum Rabi energies. Our analysis shows that only using the full Hamiltonian of Eq. \ref{eq1}, including both antiresonant terms and diamagnetic terms, it is possible to have a good fit.

In order to prove that our intersubband microcavity is indeed in the ultra-strong coupling regime, 
in the panel (a) of Fig.~\ref{fig:3} we plotted the root mean square deviation from the measured dispersion of the full Hamiltonian in Eq. \ref{eq1} (solid black line), the Hamiltonian without the anti-resonant terms (red dashed line) and the Hamiltonian without both the anti-resonant terms and the dimagnetic terms (blue dash-dotted line).
For the full Hamiltonian a perfect agreement is found for a vacuum Rabi energy $\hbar\Omega_R=\hbar\Omega_{R,k_{res}}=16.5$ meV $\sim11$~\% of the intersubband transition energy, with a fit RMS error of only 0.9 meV. 
For the other two lines the agreement is much worser, with a minimum error of $4.0$ and $7.2$ meV respectively, well beyond the experimental resolution. These minima occurs at $\hbar\Omega_{R,k_{res}}=16.5$ meV and 14.5 meV respectively.
In the bottom panels of Fig.~\ref{fig:3} the optimal angular dispersions are plotted in the three cases and compared with the experimental values (black crosses). 

Note that, as discussed in \cite{21}, the actual value of the vacuum Rabi energy is much smaller than half the splitting observed in the spectra, owing to the fact that the two polariton energies, once measured at the same angle, do not correspond to the same $k$. 

These data provide a clear demonstration that anti-resonant light-matter coupling and photon-energy renormalization can become very significant even in the interaction with the vacuum electromagnetic field of a microcavity. These anomalous contributions represent unambiguous evidence that the optoelectronic coupling is in the $ultra-strong$ regime. We believe the results show that intersubband transitions will play a key role for the development of a new quantum-optics field, thanks also to the possibility of manipulating coupling by controlling the charge density. Additionally, the fact that these phenomena can be observed at room temperature and in solid-state structures is a crucial aspect for novel device implementations.
\begin{acknowledgments}
This work was supported in part by the EC Research and Training Network POISE. We thank I.~Carusotto for useful discussions.
\end{acknowledgments}

\begin{widetext}
\begin{figure}[h!]
\begin{center}
\includegraphics*[width=\textwidth]{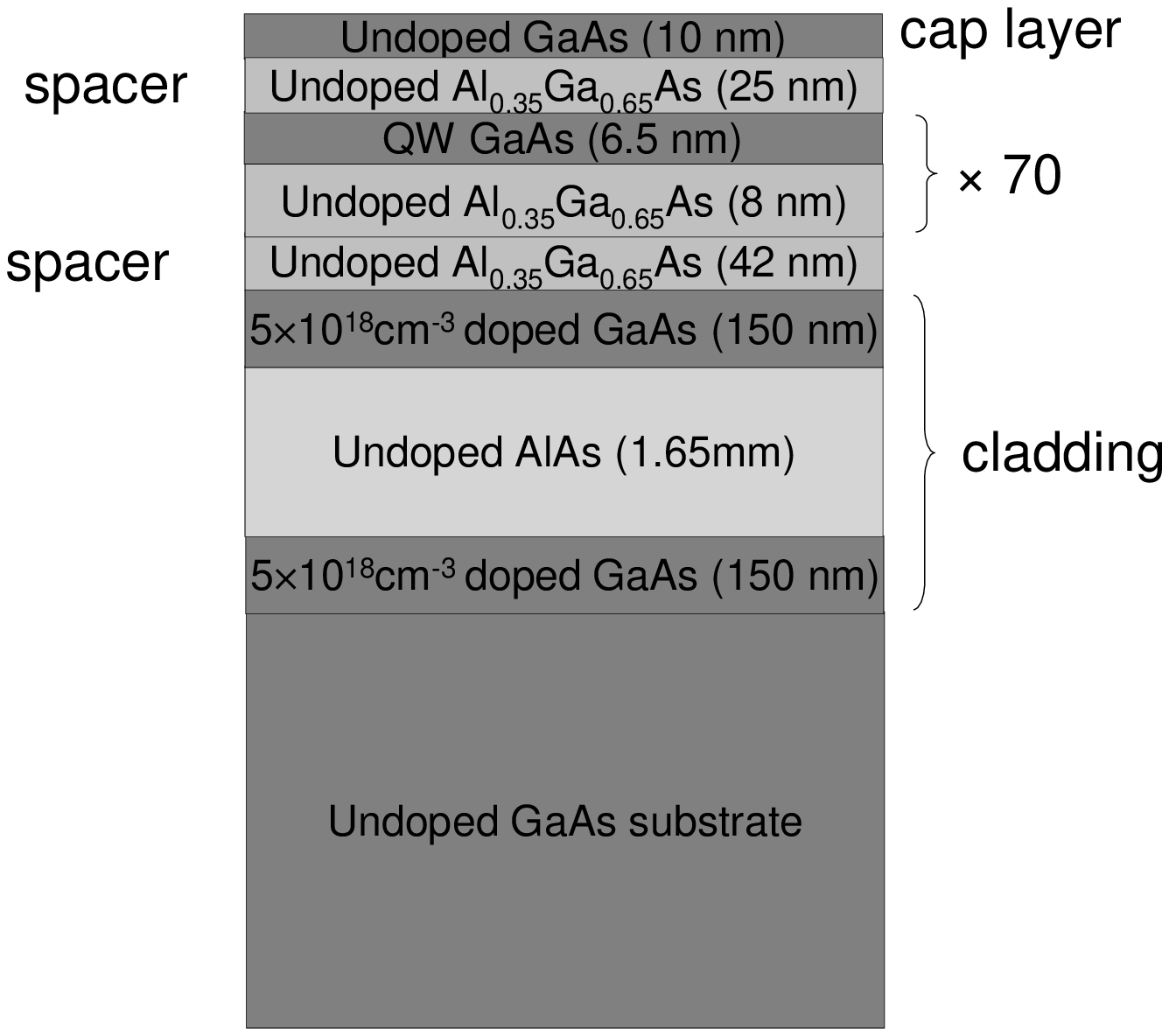}
\caption{{\bf Supplementary material.} Layer sequence of the intersubband microcavity structure}
\end{center}
\end{figure} 

\begin{figure}[h!]
\begin{center}
\includegraphics*[width=\textwidth]{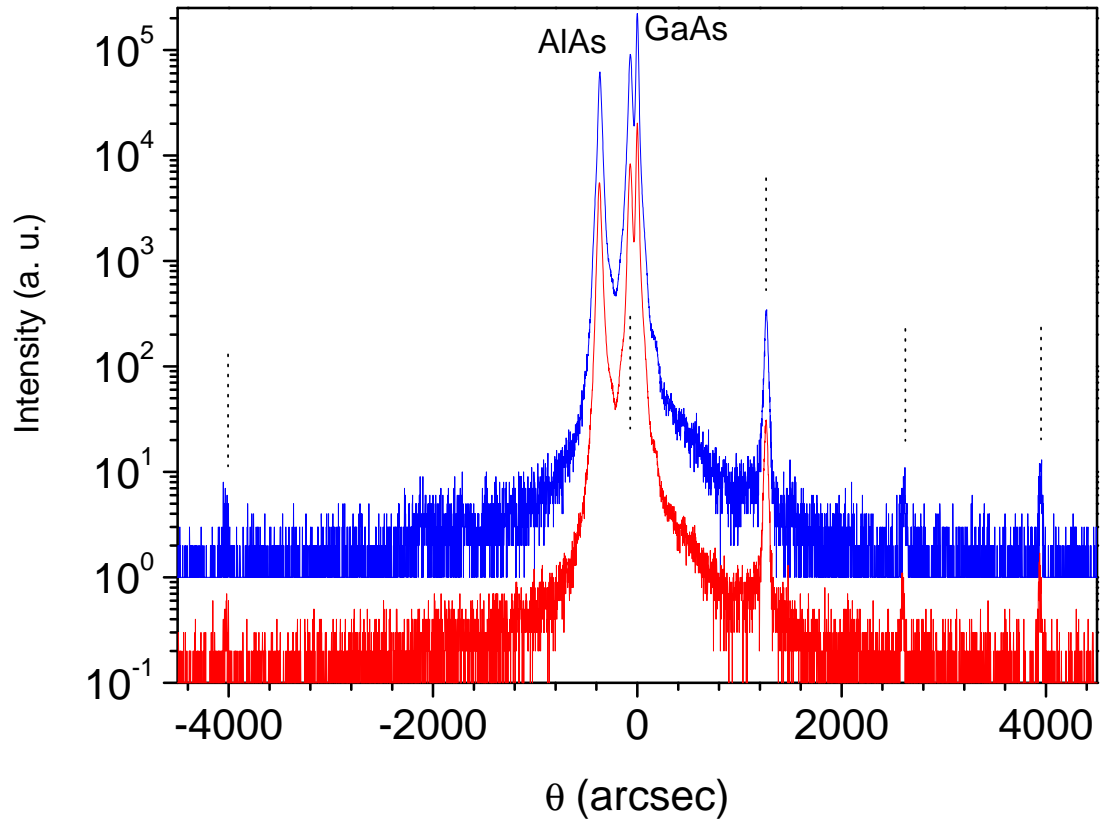}
\caption{{\bf Supplementary material.} X-ray diffraction rocking curves for the samples with doped (blue) and undoped active region (red). Curves are offset for clarity. Thickness deviations
between the two samples are below 1\%, within the measurement resolution.}
\end{center}
\end{figure} 

\begin{figure}[h!]
\begin{center}
\includegraphics*[width=\textwidth]{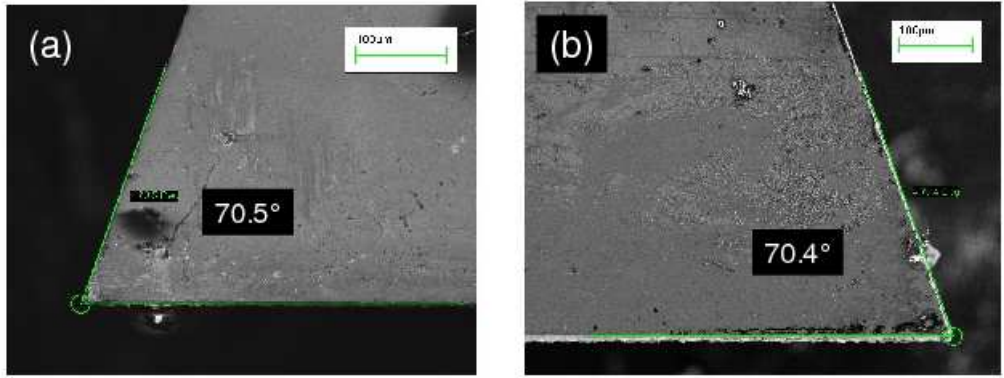}
\caption{{\bf Supplementary material.} Scanning electron microscope picture of a section of the samples wedged at 70 degrees with doped (a) and undoped (b) active region. The waveguide angles were
measured to be 70.5 degrees for the doped structure and 70.4 for the undoped one.}
\end{center}
\end{figure} 

\begin{figure}[h!]
\begin{center}
\includegraphics*[width=\textwidth]{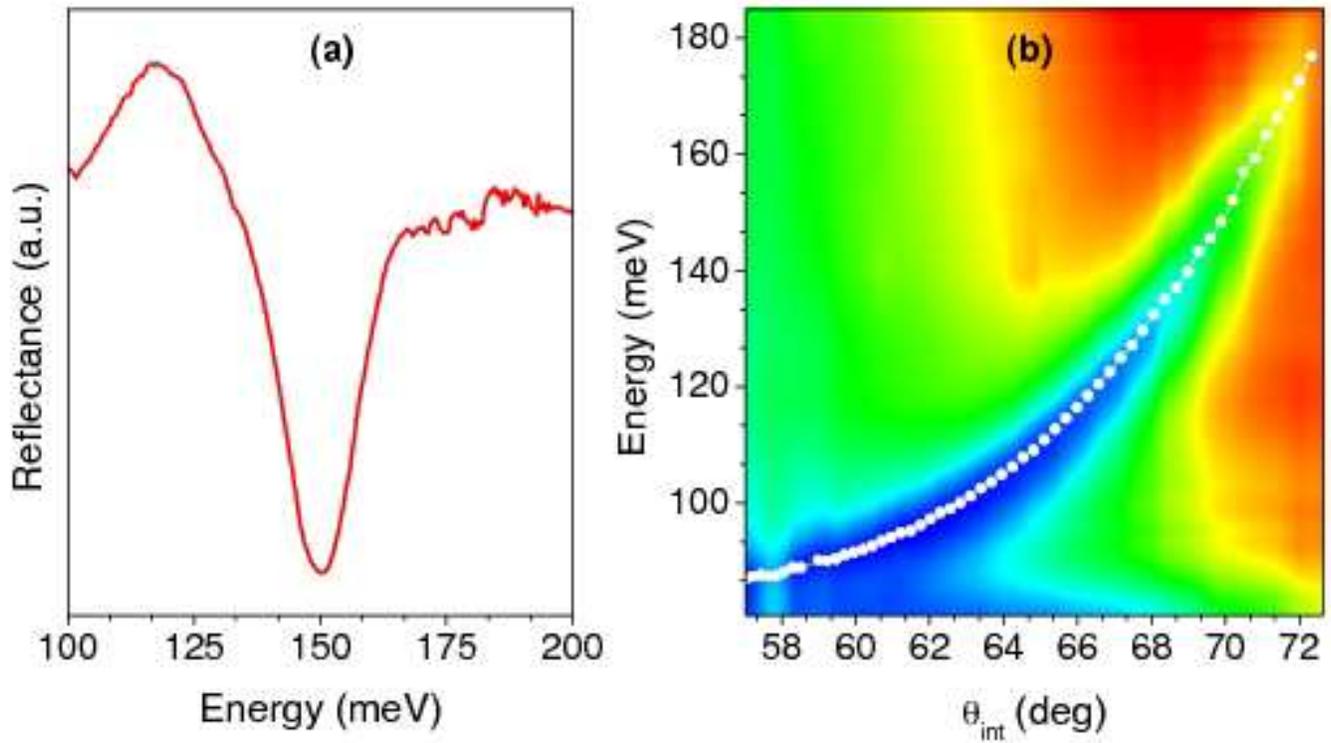}
\caption{{\bf Supplementary material.} (a) Room-temperature intersubband absorption of the quantum-well active region as measured at 37 degrees incidence angle in a sample wedged
at 45 degrees. Data was collected with a resolution of 0.25 meV. (b) Angle-resolved reflectance data of the bare-cavity sample in a 70 degree prism
displayed as contour plot (low and high reflectance values in blue and red, respectively). The measurements were done at an internal angle step of
0.3 degree, which correspond to an external angle of 1 degree. The energy position of the dip corresponding to the bare-cavity resonance is shown
as white dots.}
\end{center}
\end{figure}

\end{widetext}

\end{document}